\documentclass[12pt, slashbox]{article}
\def\BorL{b }	
\usepackage{amsmath,amssymb}
\makeatletter
\newif\ifpr@pstyle \pr@pstylefalse
\newif\ifnons@qeq  \nons@qeqfalse
\def\bigpage{
	\setlength{\topmargin}{-.5in}
	\setlength{\oddsidemargin}{.5pc}
	\setlength{\evensidemargin}{.5pc}
	\setlength{\textwidth}{35pc}
	\setlength{\textheight}{56pc}
	\setlength{\parskip}{6pt plus 2pt minus 1pt}
	\newlength{\paperbaselineskip}
	\setlength{\paperbaselineskip}{20pt plus 0.2pt minus 0.1pt}
	\def\@oddfoot{\hfil -- \thepage~--\hfil}
	\let\@evenfoot\@oddfoot
        \def\thesection{\arabic{section}.}
        \def\thesubsection{\thesection\arabic{subsection}}
        \def\@ourappendix{\par\setcounter{section}{0}
                      \setcounter{subsection}{0}
                      \def\thesection{\Alph{section}.}
                      \ifnons@qeq
                      \def\theequation{\Alph{section}.\arabic{equation}}\fi}
        \def\appendix{\@ourappendix}
        \def\section{\@startsection {section}{1}%
            {\z@}{5ex plus .2ex minus .4ex}%
            {1.5ex plus.4ex minus .1ex}%
            {\centering\ifpr@pstyle\else\ifx\undefined\reset@font\else%
             \reset@font\fi\large\fi\bf}}
        \def\subsection{\@startsection{subsection}%
            {2}{\z@}{3.25ex plus .4ex minus .4ex}%
            {1ex plus .2ex}{\bf}}
}
\bigpage
\newfont{\fourteencp}{cmcsc10 scaled\magstep2}
\newfont{\titlefont}{cmbx10 scaled\magstep2}
\newfont{\authorfont}{cmcsc10 scaled\magstep1}
\newfont{\fourteenmib}{cmmib10 scaled\magstep2}
	\skewchar\fourteenmib='177
\newfont{\elevenmib}{cmmib10 scaled\magstephalf}
	\skewchar\elevenmib='177
\newfont{\ninemib}{cmmib9} \skewchar\ninemib='177
\makeatletter
\newcommand\nonsequentialeqnum{
        \nons@qeqtrue
	\@addtoreset{equation}{section}
	\def\theequation{\arabic{section}.\arabic{equation}}}
\newif\ifp@bblock  \p@bblocktrue
\newcommand\nopubblock{\p@bblockfalse}
\newcommand\topspace{\hrule height 0pt depth 0pt \vskip}
\newcommand\p@bblock{\begingroup \tabskip=\hsize minus \hsize
	\baselineskip=1.5\ht\strutbox \topspace-2\baselineskip
	\halign to\hsize{\strut ##\hfil\tabskip=0pt\crcr
	\the\Pubnum\crcr\the\date\crcr}\endgroup}
\newcommand\YUKAWAmark{\hbox{
        \ifpr@pstyle\ninemib\else\elevenmib\fi
        Yukawa\hskip1mm Institute\hskip1mm Kyoto \hfill}}
\newtoks\date
\newtoks\Pubnum
\let\pubnum=\Pubnum
\Pubnum={}
\date={\today}
\newcommand{\frontpageskip}{\vspace{12pt plus .5fil minus 2pt}}
\def\@authoraddress{} \def\@title{}
\def\title#1{\gdef\@title{\frontpageskip
	\begin{center}{\titlefont #1}\end{center}\par}}
\def\@author#1{\frontpageskip\par\begin{center}{\authorfont #1}
	\end{center}
	\nobreak}
\def\author#1{\expandafter\def\expandafter\@authoraddress\expandafter
    {\@authoraddress{\@author{#1}}}}
\def\andauthor#1{\expandafter\def\expandafter\@authoraddress\expandafter
    {\@authoraddress{\frontpageskip\centerline{and}\@author{#1}}}}
\def\authors#1{\expandafter\def\expandafter\@authoraddress\expandafter
    {\@authoraddress{\frontpageskip\noindent #1}}}
\def\@address#1{\par\begin{center}{\sl #1}\end{center}\par}
\def\address#1{\expandafter\def\expandafter\@authoraddress\expandafter
    {\@authoraddress{\@address{#1}}}}
\def\andaddress#1{\expandafter\def\expandafter%
    \@authoraddress\expandafter
    {\@authoraddress{\par\centerline{\sl and}\@address{#1}}}}
\renewcommand{\thanks}[1]{\footnote{#1}}
\def\maketitle{\par
  \begingroup
       \def\thefootnote{\fnsymbol{footnote}}
	\thispagestyle{empty}
        \baselineskip=\paperbaselineskip
	\@maketitle
	\endgroup
	\setcounter{footnote}{0}
	\let\maketitle\relax \let\@maketitle\relax
	\let\@thanks\relax \let\@title\relax
	\let\@title\relax \let\@authoraddress\relax
	\let\thanks\relax}
\def\@maketitle{%
        \ifpr@pstyle\vspace{-1.0cm}\else\vspace{-1.7cm}\fi
	\YUKAWAmark\vskip0.6cm
	\ifp@bblock\p@bblock \else\hrule height 0pt \relax \fi
	\@title
	\@authoraddress
	}
\renewcommand{\abstract}{\par\frontpageskip\centerline{
             \ifpr@pstyle\twelvecp\else\fourteencp\fi Abstract}
	\vspace{8pt plus 3pt minus 3pt}}
\@addtoreset{equation}{section}
\def\theequation{\arabic{section}.\arabic{equation}}
\makeatother
\def\bigmode{b }
\ifx\BorL\undefined\read-1 to\BorL\fi
\makeatletter
\def\doublepage{
        \twocolumn
        
        \pr@pstyletrue
        \sloppy
        \flushbottom
        \setlength{\topmargin}{-0.95in}
        \setlength{\headsep}{20pt}
        \setlength{\headheight}{10pt}
        \hoffset=-0.35in
        \leftmargini 2em
        \leftmarginv .5em
        \leftmarginvi .5em
        \marginparwidth 48pt
        \marginparsep 10pt
        \setlength{\columnsep}{0.7truein}
        \setlength{\textwidth}{10.5truein}
        \setlength{\textheight}{7truein}
        \setlength{\oddsidemargin}{0.0truein}
        \setlength{\evensidemargin}{0.0truein}
        \multiply\paperbaselineskip by 4
                   \divide\paperbaselineskip by 5
        \multiply\footskip by 4 \divide\footskip by 5
        \setlength{\parskip}{4pt plus 1.5pt minus 1pt}
        \newlength{\halfwidth}
        \halfwidth=\textwidth\advance\halfwidth by -\columnsep
                         \divide\halfwidth by 2
        \newfont{\twelvemib}{cmmib10 scaled\magstep1}
                 \skewchar\twelvemib='177
        \newfont{\tenmib}{cmmib10}
                 \skewchar\tenmib='177
        \newfont{\twelvecp}{cmcsc10 scaled\magstep1}
        \def\pagebox{\hbox to \halfwidth{\hfil  -- \thepage~--\hfil}}
        \def\@oddfoot{\pagebox\hfil\addtocounter{page}{1}\pagebox}
        \let\@evenfoot\@oddfoot
        \def\ps@empty{\let\@mkboth\@gobbletwo\let\@oddhead\@empty
               \def\@oddfoot{\hbox to \halfwidth{\hfil ~~~~~~~}\hfil
               \addtocounter{page}{1}\pagebox}
                \let\@evenhead\@empty\let\@evenfoot\@oddfoot}
        \def\appendix{\@ourappendix}
        \def\section{\@startsection {section}{1}%
            {\z@}{5ex plus .2ex minus .4ex}%
            {1.5ex plus.4ex minus .1ex}%
            {\centering\ifpr@pstyle\else\reset@font\large\fi\bf}}
        \def\subsection{\@startsection{subsection}%
            {2}{\z@}{3.25ex plus .4ex minus .4ex}%
            {1ex plus .2ex}{\bf}}
}
\ifx\BorL\bigmode
\else
        \input art10.sty
        \doublepage
\fi
\makeatother
\makeatletter
\newif\ifepsfloaded
\newif\iffigureexists

\ifeof 1 \epsfloadedfalse \else \epsfloadedtrue \fi
\ifepsfloaded \input epsf \fi

\def\checkex#1 {\relax
    \openin 1 #1
    \ifeof 1 \figureexistsfalse
    \else \figureexiststrue
    \fi \closein 1 }
\def\figinsertraw#1#2{
   \ifepsfloaded
       \checkex #1
       \iffigureexists
           \immediate\write16{(#1)}
           #2
       \else
           \immediate\write16{(#1 NOT FOUND!)}
           \vbox to 2in{\hbox to 2in {\hss} \vss}
       \fi
   \else
       \immediate\write16{(NOT inputting #1; no epsf.tex)}
       \vbox to 2in{\hbox to 2in {\hss} \vss}
   \fi}
\newcommand{\reduceland}[2]{\dimen@=#1
     \ifpr@pstyle\multiply\dimen@ by 4\divide\dimen@ by 5\fi
     \edef#2{\dimen@}}
\def\F@gin#1#2#3#4{
  \ifepsfloaded
    \checkex #1
    \iffigureexists
        \immediate\write16{(#1)}
        \begin{figure}
        \ifdim#2>\z@\reduceland{#2}{\dimen@ii}\epsfxsize=\dimen@ii\fi
        \ifdim#3>\z@\reduceland{#3}{\dimen@ii}\epsfysize=\dimen@ii\fi
        \centerline{\epsfbox{#1}}
        {#4} \end{figure}
    \else
        \immediate\write16{(#1 NOT FOUND!)}
        \begin{figure}
        \ifdim#2>\z@\reduceland{#2}{\dimen@ii}\else\dimen@ii=2in\fi
        \ifdim#3>\z@\reduceland{#3}{\dimen255}\else\dimen255=2in\fi
        \centerline{\framebox[\dimen@ii]{\rule{0pt}{\dimen255}#1}}
        {#4} \end{figure}
    \fi
  \else
    \immediate\write16{(NOT inputting #1; no epsf.tex)}
    \begin{figure}
    \centerline{\framebox[2in]{\rule{0pt}{2in}#1}}
    #4\end{figure}
  \fi}
\def\figinsertx#1#2#3{\F@gin{#1}{#2}{0pt}{#3}}
\def\figinserty#1#2#3{\F@gin{#1}{0pt}{#2}{#3}}
\def\figinsert#1#2{\F@gin{#1}{0pt}{0pt}{#2}}
\makeatother
\begin{document}    
%
\renewcommand{\thefootnote}{\fnsymbol{footnote}}
\pubnum{YITP-04-57\cr OIQP-04-4}
\date{Oct. 2004}

\title{Dirac Sea for Bosons II\\
--- Study of the Naive Vacuum Theory for the Playground World Prior to
    Filling the Negative Energy Sea ---
\thanks{This paper is the second part of the revised version of ref. [2].}
}

\author{Holger B. Nielsen
}
\address{Niels Bohr Institute, \\
	University of Copenhagen, 17 Blegdamsvej, \\
	Copenhagen $\o$, DK 2100, Denmark\\
}

\andauthor{Masao Ninomiya
\thanks{Also, Okayama Institute for Quantum Physics, Kyo-yama 1-9-1,
Okayama City 700-0015, Japan 
}
}
\address{Yukawa Institute for Theoretical Physics\\
        Kyoto University,~Sakyo-ku,~Kyoto 606-8502,~Japan\\}

\maketitle


\begin{abstract}
We use our previous idea[1] in the foregoing paper entitled Dirac Sea for
Bosons I, in which at first we perform a naive second
quantization of both negative and positive energy for the Klein-Gordon
equation analogous to the unfilled Dirac sea for fermions, in order to study as
a playground this naive second quantization theory.
It is not to be taken physically satisfactory model in as far as it has
Fock space with indefinite signature of norm square, but it has
nevertheless interesting possibilities: Although the naive
(quantization) theory represents a spontaneous breakdown of the usual
CPT-symmetry, we shall show that it obeys a certain replacement called
a CPT-symmetry for the CPT-theorem for which properties and a proof are
presented.

\end{abstract}


\section{Introduction}
Dirac's picture of a vacuum state in which negative energy solutions
of the Dirac equation has in our opinion a lot of intuitive appeal
although nowadays one often meets it in so formal dressing that the
beauty of the picture disappeared.
One just only writes the second quantized fields as a
superposition of both creation and annihilation operators, namely
creation of antiparticles and annihilation of particles in say $\mathbf{\Psi}(x)$.

Of course at the end it does not matter what way one formulates the
second quantized theory, but as an intermediate step one has in the
viewpoint of the Dirac sea the picture of a somewhat different theory,
namely the theory in which the vacuum has {\em not yet been filled.}
Of course this theory before the Dirac sea is filled is not good
physically in as far as it has no bottom in the Hamiltonian. 
It is thus clearly not realistic physically, but it is in certain
aspects simpler and as a theoretical playground it may have some
attractive features. 

In the foregoing paper ``Dirac Sea for Bosons I''[1] in this series we
proposed how one could also formally consider Dirac sea picture for
bosons.
We only there succeeded in doing that by using with {\em negative}
numbers of bosons.
But using the trick of playing with allowing the negative number of
bosons in a certain state/orbit 
we managed to formulate the physically true second quantized
theory as the result of second quantizing the boson in a naive way, in
which the second quantized field is expanded only on annihilation
operators (but no creation ones).
Next we then performed a modification of the theory analogous to the
filling of the Dirac sea for fermion case.
This operation was then not a filling but rather a passage of the
barrier -- as we called it -- between the particle numbers 0, 1, 2,
... and thenumbers -1, -2, -3, ... in the
negative energy single particle states.

Really this means that we first quantized ``naively'' with the second
quantized field $\mathbf{\Psi}$ written as an expansion on annihilation
operators alone.
Then afterwards we switched across the barrier so as to have in
all the negative energy states/orbit -1 boson.
This then meant that one could in these state only {\em remove} bosons
but not add any.
Such a removal -- a kind of hole -- was then interpreted as the
antiparitcle.

In the present paper, we intend to study the theory \underline{before}
the filling of
the Dirac sea and the analogous operation for the bosons.
In the long run it is our hope to develop a way: first consider this
in many ways attractive theory and then taking into account later the
effect of the Dirac sea and of the analogous modification of the
boson vacuum.  
In the present article we shall discuss some of the beauties of
this ``pre-filling'' theory in spite of its severe physical
drawbacks.

Some of the advantages of the theory of the naive quantization 
is that for a finite number of n particles
{\em on top of} the naive vacuum world.
One can describe the
whole theory by a wave function describing just n particles.
In the case of the filled Dirac sea the wave function in which one can
have the particles must be orthogonal to the negative energy states,
since otherwise it would be modified when the antisymmetrization of
the total wave function would be performed including the sea.
Indeed an n-particle state in addition to, so to speak, the normal Dirac
sea (for fermions) would have to have its single particle states
orthogonal to the negative energy states, which means they should be
pure positive energy states.
One might consider such requirements as a complication; it could be
nice to avoid it in order to have some simpler thinking on some other
aspects of the considered situation.
This could be a motivation for, for instance, hope for understanding
some bound state phenomena better using an emptied
sea.\footnote{Akierzer and Berestetski dreamed such phenomena in a atom
using Dirac equation [3].}

If all interactions were really only infinitely short range the wave
functions for scattering would be composed from very simple
contributions and the theory basically solvable.
In the example of an interacting bosonic theory in the naive vacuum
treated below we use the interaction which in the second
quantized formulation take the form of a term 

\begin{equation}
L_{int}=\frac{\lambda}{4}{\left(\pmb{\Psi}^+\right)}^2 
\pmb{\Psi}^2+h.c.
\label{1.1}
\end{equation}

Thinking of the boson as charged with a ``particle number'' this
charge is conserved in the interaction eq.(\ref{1.1}) and so this
interaction only makes the particles scatter but does not change their
particle number. 
Not having the Dirac sea the number of particles would therefore
stay finite even with this interaction eq.(\ref{1.1}) and the same value if
it starts finite.
Having the sea would always be more complicated because we then really
have infinitely many particles.

In the present article this simplified world of finite numbers of
particles in the naive vacuum is put forward with the just mentioned
example eq.(1.1), and as one of the points we are doing with it is a kind of
replacement of CPT-symmetry.
This is actually, as we shall see below, a quite different symmetry from
the usual CPT-symmetry[4].
The usual one would let the naive vacuum go into a quite different
vacuum and thus would rather connect theories built on different vacua
than being a symmetry inside such a single theory.

In the following section 2 we present the scalar bosonic theory with
the interaction eq.(\ref{1.1}) and especially describe the scattering
between the particles in terms of phase shifts.
In section 3 we show that some analyticity properties result in the
case of scattering of particles coming in from infinity like in
S-matrix construction.
In fact we even study the analytical continuation of these wave
functions as analytical functions and continue them into other sheets.

Then in section 4 we formulate, as our replacement of usual
CPT-theorem, CPT-like theorem. 
In section 5 we put forward a proof or at
least indication for a proof
of this new CPT-like theorem for the naive vacuum theory.
In section 6 we discuss relation between CPT theorem and to the usual
CPT-theorem,
and in section 7 we give the review and the conclusion.

\section{The $(\varphi^+\varphi)^2$ theory and boundary condition}

In this section we illustrate how to rewrite an interaction into a
boundary condition for the wave function by taking as an example the
scattering in the $\lambda(\varphi^+\varphi)^2$ theory.
The field expansion of the second quantized field $\hat\varphi$ reads

\begin{equation}
\hat\varphi(x)=\sum_{\stackrel{\rightharpoonup}{k},\pm}\left\{
\frac{1}{\sqrt{2\omega}}
e^{-i\omega t+i\stackrel{\rightharpoonup}{k}\cdot\stackrel{\rightharpoonup}{x}}
\cdot a(\stackrel{\rightharpoonup}{k},+) +\frac{1}{\sqrt{2\omega}}
e^{i\omega t +i\stackrel{\rightharpoonup}{k}\cdot\stackrel{\rightharpoonup}{x}}
\cdot a(\stackrel{\rightharpoonup}{k},-)\right\}
\end{equation}
and we insert it into the expression for the interaction Hamiltonian 

\begin{equation}
H_{int}=\frac{\lambda}{4}\int
d^3\stackrel{\rightharpoonup}{x}\hat\varphi^+(\stackrel{\rightharpoonup}{x},t)^2\hat\varphi 
(\stackrel{\rightharpoonup}{x},t)^2 \ \ \ .
\end{equation}
With this interaction there is only the s-wave scattering in the
partial wave analysis for two charged scalars.
In the partial wave analysis, the delay or advance of the
scattered particles is given by $\frac{d\delta(\omega)}{d\omega}$ , 
where $\delta(\omega)$ is the energy of one of the outgoing particles.
Thus we may write for very short range interaction 

\begin{equation}
\frac{d\delta(\omega)}{d\omega}=0
\end{equation}
In terms of the spatial wave function in the s-wave channel, the phase 
shift is defined such that the s-wave wave function should be of the
form 
\begin{equation}
\cos\delta(\omega)\cdot j_0(k\cdot r)-
\sin\delta(\omega)\cdot n_0(k\cdot r) \ \ \  .
\label{7.4}
\end{equation}
Here $r$ is the relative distance in the center of mass system and
$\stackrel{\rightharpoonup}{k}$ the relative momentum.
The functions $j_0(k\cdot r)$ and $n_0(k\cdot r)$ denote

\begin{equation}
j_0(k\cdot r) = \frac{\sin k\cdot r}{k\cdot r}
\end{equation}
and 
\begin{equation}
n_0(k\cdot r) = -\frac{\cos k\cdot r}{k\cdot r} \ \ \ .
\end{equation}
It should be noticed that the four momentum in the center of mass
system can be space-like in our naive vacuum and 
thus, strictly speaking, a proper center of mass system does not exist.
However, we may not investigate that case furthermore.

The wave function is expressed in a superposition of eq.(\ref{7.4}) as

\begin{equation}
\varphi(\stackrel{\rightharpoonup}{x}_{rel},\stackrel{\rightharpoonup}{x})=
\int c(\omega) e^{-2i\omega t}\left\{\cos\delta (\omega)j_0(k\cdot r)-
\sin\delta(\omega)n_0(k\cdot r)\right\}d\omega
\end{equation}
where $t$ is the time in the center of mass system.

For very short relative distances we have as $r \rightarrow 0$,

\begin{eqnarray}
n_0(k\cdot r)&\rightarrow&-\frac{1}{kr}+O(k\cdot r) \ \ \ ,\nonumber\\
j_0(k\cdot r)&\rightarrow&  \ 1 \ \ .
\end{eqnarray}
We can extract the coefficients of $n_0$ and $j_0$ as the ones to the 
$\frac{1}{r}$-part and constant part of $\varphi$, which can be
estimated as 

\begin{eqnarray}
{\rm coeff. \  of \ } 
\varphi|_{\frac{1}{r}-part}&=&\lim _{r\rightarrow 0}r\varphi\nonumber\\
&=&\int c(\omega)\sin\delta(\omega)\frac{e^{-2i\omega t}}{k}
d\omega
\end{eqnarray}
and
\begin{eqnarray}
{\rm coeff. \  of \ } 
\varphi|_{const-part}&=&\lim_{r\rightarrow
0}(\varphi-\frac{1}{r}\varphi|{}_{\frac{1}{r}-part})\nonumber\\
&=&\int c(\omega)\cos\delta(\omega)e^{-2i\omega t}d\omega \ \ \ .
\end{eqnarray}
For large $|\omega|$ the masses may be ignored and $\omega$ behaves as 

\begin{equation}
|\omega| \sim k \ \ \ .
\end{equation}
By taking $\delta(\omega)=\delta$ which is constant we find a relation 

\begin{equation}
i
\left({\rm coeff. \ of \ }\dot\varphi|_{\frac{1}{r}-part}\right)=
	2\tan \delta\cdot\left({\rm coeff. \ of \ }
	\varphi|_{const-part}\right)\ \ \ .
\label{7.12}
\end{equation}
We then evaluate how the constant phase shift $\delta$ is related to
the coupling constant $\lambda$.

Before going into this evaluation we must construct a wave function
for two particle state by using the $(\pi,\varphi$)-formalism where
$\pi$ denotes conjugate momentum to $\varphi$.
An $N$-scalar wave function has $2^N$ components and each of those
are marked as $\varphi$ or $\pi$ with respect to each of the $N$
particles in the state.
For example a two particle state wave function may be written as 

\begin{equation}
\left(
\begin{array}{cc}
\varphi_{\varphi\varphi}&
	\hskip-2\arraycolsep(\stackrel{\rightharpoonup}{x_1},
	\stackrel{\rightharpoonup}{x_2}; t)\\
\varphi_{\varphi\pi}&
	\hskip-2\arraycolsep(\stackrel{\rightharpoonup}{x_1},
	\stackrel{\rightharpoonup}{x_2}; t)\\
\varphi_{\pi\varphi}&
	\hskip-2\arraycolsep(\stackrel{\rightharpoonup}{x_1},
	\stackrel{\rightharpoonup}{x_2}; t)\\
\varphi_{\pi\pi}&
	\hskip-2\arraycolsep(\stackrel{\rightharpoonup}{x_1},
	\stackrel{\rightharpoonup}{x_2}; t)
\end{array}
\right) \ \ \ .
\end{equation}
The inner product of these wave functions can be defined by extending
the one for single particle states.
It reads

\begin{equation}
\left<\left(
\begin{array}{l}
\varphi_\varphi\\
\varphi_\pi
\end{array}
\right)_1\Bigg|
\left(
\begin{array}{l}
\varphi_\varphi\\
\varphi_\pi
\end{array}
\right)_2\right>=
-i\int(\varphi^*_{\pi 1}\varphi_{\varphi 2}-
\varphi^*_{\varphi1}\varphi_{\pi 2}) \ d^3 \stackrel{\rightharpoonup}{x}
\end{equation}
where the subscript 1 and 2 denote two single particle states.
This expression may be extended to the two particle system as 

\begin{equation}
\left<\left(
\begin{array}{l}\varphi_{\varphi\varphi1}\\
		\varphi_{\varphi\pi1}\\
		\varphi_{\pi\varphi1}\\
		\varphi_{\pi\pi1}
\end{array}\right)\Bigg|\left(
\begin{array}{l}\varphi_{\varphi\varphi2}\\
		\varphi_{\varphi_\pi2}\\
		\varphi_{\pi\varphi2}\\
		\varphi_{\pi\pi2}
\end{array}\right)\right>
=\int(\varphi^*_{\varphi\varphi}
	\varphi^*_{\varphi\pi}
	\varphi^*_{\pi\varphi}
	\varphi^*_{\pi\pi})_1
\left(
\begin{array}{cccc}
0&0&0&-1\\
0&0&1&0\\
0&1&0&0\\
-1&0&0&0
\end{array}
\right)
\left(
\begin{array}{l}\varphi_{\varphi\varphi}\\
		\varphi_{\varphi\pi}\\
		\varphi_{\pi\varphi}\\
		\varphi_{\pi\pi}
\end{array}\right)_2
d^3\stackrel{\rightharpoonup}{x_1} \
d^3\stackrel{\rightharpoonup}{x_2} \ \ \ .
\end{equation}

The second quantized filed
$\hat\varphi(\stackrel{\rightharpoonup}{x})$ 
acts on the two particle state as 

\begin{equation}
\hat\varphi(\stackrel{\rightharpoonup}{x})
\left(
\begin{array}{l}
\varphi_{\varphi\varphi}
	(\stackrel{\rightharpoonup}{x_1}\ , \ 
	 \stackrel{\rightharpoonup}{x_2})\\
\varphi_{\varphi\pi}
	(\stackrel{\rightharpoonup}{x_1}\ , \ 
	 \stackrel{\rightharpoonup}{x_2})\\
\varphi_{\pi\varphi}
	(\stackrel{\rightharpoonup}{x_1}\ , \ 
	 \stackrel{\rightharpoonup}{x_2})\\
\varphi_{\pi\pi}
	(\stackrel{\rightharpoonup}{x_1}\ , \ 
	 \stackrel{\rightharpoonup}{x_2})
\end{array}\right)=\left(
\begin{array}{l}
\varphi_\varphi
	(\stackrel{\rightharpoonup}{x_1})\\
\varphi_\pi
	(\stackrel{\rightharpoonup}{x_1})
\end{array}
\right) 
\delta^3(\stackrel{\rightharpoonup}{x}-
	\stackrel{\rightharpoonup}{x_2})+\left(
\begin{array}{l}
\varphi_\varphi
	(\stackrel{\rightharpoonup}{x_2})\\
\varphi_\pi
	(\stackrel{\rightharpoonup}{x_2})
\end{array}
\right)
\delta^3(\stackrel{\rightharpoonup}{x}-
	\stackrel{\rightharpoonup}{x_1})
\ \ \ .
\end{equation}

Thus $\hat\varphi(\stackrel{\rightharpoonup}{x})$ is just a pure
annihilation operator.
On the other hand, the $\hat\pi$-operator acts as

\begin{equation}
\hat\pi(\stackrel{\rightharpoonup}{x})
\left(
\begin{array}{l}
\varphi_{\varphi\varphi}
	(\stackrel{\rightharpoonup}{x_1}\ , \ 
	 \stackrel{\rightharpoonup}{x_2})\\
\varphi_{\varphi\pi}
	(\stackrel{\rightharpoonup}{x_1}\ , \ 
	 \stackrel{\rightharpoonup}{x_2})\\
\varphi_{\pi\varphi}
	(\stackrel{\rightharpoonup}{x_1}\ , \ 
	 \stackrel{\rightharpoonup}{x_2})\\
\varphi_{\pi\pi}
	(\stackrel{\rightharpoonup}{x_1}\ , \ 
	 \stackrel{\rightharpoonup}{x_2})
\end{array}\right)=\left(
\begin{array}{l}
\varphi_\varphi
	(\stackrel{\rightharpoonup}{x_1})\\
\varphi_\pi
	(\stackrel{\rightharpoonup}{x_1})
\end{array}
\right) 
\delta^3(\stackrel{\rightharpoonup}{x}-
	\stackrel{\rightharpoonup}{x_2})+\left(
\begin{array}{l}
\varphi_\varphi
	(\stackrel{\rightharpoonup}{x_2})\\
\varphi_\pi
	(\stackrel{\rightharpoonup}{x_2})
\end{array}
\right)
\delta^3(\stackrel{\rightharpoonup}{x}-
	\stackrel{\rightharpoonup}{x_1})
\ \ \ .
\end{equation}

Next we consider the Hermitean conjugate operator 
$\hat\varphi(\stackrel{\rightharpoonup}{x})^+$
which should act on the states in such a way that it has the same 
effect as $\varphi(\stackrel{\rightharpoonup}{x})$ in the ket.
Thus

\begin{equation}
\hat\varphi^+(\stackrel{\rightharpoonup}{x})
\left(
\begin{array}{l}
	\varphi_\varphi(\stackrel{\rightharpoonup}{x_1})\\
	\varphi_\pi(\stackrel{\rightharpoonup}{x_1})
\end{array}\right)=\left(
\begin{array}{lcl}
\varphi_{\varphi\varphi}
	(\stackrel{\rightharpoonup}{x_1}\ , \ 
	 \stackrel{\rightharpoonup}{x_2})&=&0\\
\varphi_{\varphi\pi}
	(\stackrel{\rightharpoonup}{x_1}\ , \ 
	 \stackrel{\rightharpoonup}{x_2})&=&
	\varphi_\varphi(\stackrel{\rightharpoonup}{x_1})\delta^3
	(\stackrel{\rightharpoonup}{x_2}-\stackrel{\rightharpoonup}{x})\\
\varphi_{\pi\varphi}
	(\stackrel{\rightharpoonup}{x_1}\ , \ 
	 \stackrel{\rightharpoonup}{x_2})&=&0\\
\varphi_{\pi\pi}
	(\stackrel{\rightharpoonup}{x_1}\ , \ 
	 \stackrel{\rightharpoonup}{x_2})&=&
	\varphi_\pi(\stackrel{\rightharpoonup}{x_1})\delta^3
	(\stackrel{\rightharpoonup}{x_2}-\stackrel{\rightharpoonup}{x})
\end{array}\right)
\end{equation}

Therefore starting from zero particle state $\varphi_0$ we obtain

\begin{eqnarray}
\left(\hat\varphi^+
	\left(\stackrel{\rightharpoonup}{x}\right)\right)^2\varphi_0&=&
	\hat\varphi(\stackrel{\rightharpoonup}{x})^+
	\left(
\begin{array}{l}
	\varphi_\varphi(\stackrel{\rightharpoonup}{x_1})=0\\
	\varphi_\pi(\stackrel{\rightharpoonup}{x_1})=
		\delta^3(\stackrel{\rightharpoonup}{x}-x_1)
\end{array}\right)\nonumber\\
&=&\left(
\begin{array}{l}
	\varphi_{\varphi\varphi}
	(\stackrel{\rightharpoonup}{x_1} \ , \
 	\stackrel{\rightharpoonup}{x_2})=0\\
	\varphi_{\varphi\pi}
(\stackrel{\rightharpoonup}{x_1} \ , \
 	\stackrel{\rightharpoonup}{x_2})=0\\
	\varphi_{\pi\varphi}
	(\stackrel{\rightharpoonup}{x_1} \ , \
 	\stackrel{\rightharpoonup}{x_2})=0\\
	\varphi_{\pi\pi}
	(\stackrel{\rightharpoonup}{x_1} \ , \
 	\stackrel{\rightharpoonup}{x_2})=
	\delta^3(\stackrel{\rightharpoonup}{x}-
	\stackrel{\rightharpoonup}{x_1})
	\delta^3(\stackrel{\rightharpoonup}{x}-
	\stackrel{\rightharpoonup}{x_2})
\end{array}\right)
\end{eqnarray}

It is now clear how the Hamiltonian operator $\hat H$ should be 
expressed by the second quantized fields 
$\hat\varphi$ and $\hat\varphi^+$ .
The free Hamiltonian operator 

\begin{equation}
\hat H_{free}=\int\left\{
	\pi^+(\stackrel{\rightharpoonup}{x})
	\pi(x)+
	(\bigtriangledown\varphi^+(\stackrel{\rightharpoonup}{x}))\cdot
	(\bigtriangledown\varphi(\stackrel{\rightharpoonup}{x}))\right\}
	d^3\stackrel{\rightharpoonup}{x}
\end{equation}

acts on the two particle states as 

\begin{equation}
\hat H_{free}\left(
\begin{array}{l}
\varphi_{\varphi\varphi}\\
\varphi_{\varphi\pi}\\
\varphi_{\pi\varphi}\\
\varphi_{\pi\pi}
\end{array}\right)=\left(
\begin{array}{llll}
0&i&i&0\\
-i\Delta&0&0&i\\
-i\Delta&0&0&i\\
0&-i\Delta&-i\Delta&0
\end{array}\right)\left(
\begin{array}{l}
\varphi_{\varphi\varphi}\\
\varphi_{\varphi\pi}\\
\varphi_{\pi\varphi}\\
\varphi_{\pi\pi}
\end{array}\right) \ \ \ .
\end{equation}

For the interaction Hamiltonian 

\begin{equation}
\hat H_{int}=\frac{\lambda}{8}\int
(\varphi(\stackrel{\rightharpoonup}{x})^+)^2
\varphi(\stackrel{\rightharpoonup}{x})^2 d^3\stackrel{\rightharpoonup}{x}
\end{equation}

we have 

\begin{equation}
\hat H_{int}\left(
\begin{array}{l}
\varphi_{\varphi\varphi}
	(\stackrel{\rightharpoonup}{x_1},
		\stackrel{\rightharpoonup}{x_2} ; t)\\
\varphi_{\varphi\pi}
	(\stackrel{\rightharpoonup}{x_1},
		\stackrel{\rightharpoonup}{x_2} ; t)\\
\varphi_{\pi\varphi}
	(\stackrel{\rightharpoonup}{x_1},
		\stackrel{\rightharpoonup}{x_2} ; t)\\
\varphi_{\pi\pi}
	(\stackrel{\rightharpoonup}{x_1},
		\stackrel{\rightharpoonup}{x_2} ; t)
\end{array}\right)=\frac{\lambda}{8}\left(
\begin{array}{l}
0\\
0\\
0\\
\delta^3(\stackrel{\rightharpoonup}{x_1}-\stackrel{\rightharpoonup}{x_2})
	\varphi_{\varphi\varphi}
	(\stackrel{\rightharpoonup}{x_1},
		\stackrel{\rightharpoonup}{x_2} ; t)
\end{array}\right) \ \ \ .
\end{equation}

In an energy eigenstate the average of $\hat H_{int}$ is given as 

\begin{eqnarray}
\left<\hat H_{int}\right>&=&\left<\left(
\begin{array}{l}
\varphi_{\varphi\varphi}\\
\varphi_{\pi\pi}\\
\varphi_{\pi\varphi}\\
\varphi_{\pi\pi}
\end{array}\right)\Bigg|\hat H_{int}\Bigg|\left(
\begin{array}{l}
\varphi_{\varphi\varphi}\\
\varphi_{\varphi\pi}\\
\varphi_{\pi\varphi}\\
\varphi_{\pi\pi}
\end{array}\right)\right> \Bigg/ \left<\left(
\begin{array}{l}
\varphi_{\varphi\varphi}\\
\varphi_{\varphi\pi}\\
\varphi_{\pi\varphi}\\
\varphi_{\pi\pi}
\end{array}\right)\Bigg|\left(
\begin{array}{l}
\varphi_{\varphi\varphi}\\
\varphi_{\varphi\pi}\\
\varphi_{\pi\varphi}\\
\varphi_{\pi\pi}
\end{array}\right)\right>\nonumber\\
&=&<\frac{\lambda}{8\omega^2}\delta^3
(\stackrel{\rightharpoonup}{x_1}-\stackrel{\rightharpoonup}{x_2})>
\end{eqnarray}

where $<>$ denotes to take the expectation value in terms of  the eigenfunction
of $\hat H_{free}$ with the eigenvalue $\omega$

In order to estimate the phase shift caused by the interaction 
$\frac{\lambda}{8}(\varphi^+\varphi)^2$
we consider an incoming wave packet with zero angular momentum
corresponding to the $s$-wave, which has a distance $L$ in the radial
direction .
It is expressed by 
$
ch_0(k\cdot r) $
as 
$ h_0(k\cdot r)\sim\frac{e^{-ik\cdot r}}{k\cdot r}$
where a coefficient $c$ is determined by 

\begin{equation}
\int^{R+\frac{L}{2}}_{R-\frac{L}{2}} |
ch_0(k\cdot r)|^2 \ \cdot \ 4\pi r^2dr=\frac{1}{(2\omega)^2} \ \ \ .
\end{equation}

Here $R$ denotes the radial position of the center
and we use normalization so that after dropping center of mass motion
it is given by 
$-i\int(\pi^*\varphi-\varphi^*\pi)d^3\stackrel{\rightharpoonup}{x}=1$.
For $\omega=k$ it gives

\begin{equation}
c=\frac{1}{4\sqrt{\pi L}} \ \ \ .
\end{equation}
while the wave passes with time $L$ the square of the wave function 
$|\varphi_{\varphi\varphi}|^2$ is given by 

\begin{equation}
|cj_0|^2=|c|^2=\frac{1}{16\pi L} \ \ \ .
\end{equation}
Thus the expectation value of $\hat H_{int}$
in the period $L$ is evaluated as 

\begin{equation}
\frac{\lambda}{8}
\int\varphi^*_{\varphi\varphi}\varphi_{\varphi\varphi}
d^3\stackrel{\rightharpoonup}{x}=
\frac{\lambda}{8}|c|^2=
\frac{\lambda}{128\pi L}
\end{equation}
and so the phase shift of the wave function on the average is given by 

\begin{equation}
\delta=-\frac{\lambda}{128\pi L}L=-\frac{\lambda}{128\pi} \ \ \ .
\end{equation}
Using this relation to the boundary condition eq.(\ref{7.12}), 
it is expressed in terms of $\lambda$ 

\begin{equation}
{\rm coeff. \ of \ }
\dot\varphi_{\varphi\varphi}|_{\frac{1}{r}-part}=-
\frac{\lambda}{64\pi}
({\rm coeff. \ of \ }\varphi_{\varphi\varphi}|_{const-part})
\label{7.30}
\end{equation}
where small $\delta$ approximation is used.
Furthermore if one uses the following relation in the center of mass frame 

\begin{equation}
\dot\varphi_{\varphi\varphi}=\varphi_{\varphi\pi}=\varphi_{\pi\varphi}
\end{equation}
we can write the boundary condition  eq.(\ref{7.30}) as

\begin{equation}
{\rm coeff. \ of \ }
\varphi_{\varphi\pi}|_{\frac{1}{r}-part}=-
\frac{\lambda}{64\pi}\cdot
({\rm coeff. \ of \ }\varphi_{\varphi\varphi}|_{const-part}) \ \ \ .
\end{equation}

In general one may rewrite local interactions as boundary conditions
in the similar manner to the above example.

\section{The analyticity and sheet structure}

We argue in this section that it is natural to think of analytic wave
functions in the naive vacuum world.
The only exception should be branch-point singularities 
where one of the distances 
$r_{ik}=
	\sqrt{(\stackrel{\rightharpoonup}{x_i}-
		\stackrel{\rightharpoonup}{x_k})^2}$ 
between a couple of particle vanishes.

However, as will be shown below the scattering of negative energy
particles on positive energy ones leads to bad singularities in the
wave functions.
To argue for these singularities we start from scattering of two particles 
with ordinary time-like total four momentum $p^\mu_1+p^\mu_2$ .
The local interaction, typically in $s$-waves, leads to a scattering
amplitude expressed by $\eta_0(kr)=-\frac{\cos kr}{kr}$ which 
is singular at $k\cdot r=0$.
Here $k$ and $r$ denote the relative momentum and distance.
It has a pole in $k\cdot r$ but has a square root branch point as a
function of 
$(kr)^2\propto[(p_1+p_2)(x_1-x_2)]^2$.
Then we analytically continue $p^\mu_1+p^\mu_2$ into space-like one.
This is what one can achieve in the naive vacuum world.
When the two formal times $x^0_1$ and $x^0_2$ are equal $x^0_1=x^0_2$
we have the true wave function.
Even under this restriction $x^0_1=x^0_2$ , in an appropriate
reference frame 
we can uphold the zero of $k\cdot r$ and thus branch point
singularities for some real $x^\mu_1-x^\mu_2$ and $p^\mu_1+p^\mu_2$.

One would be worried that such singularities might challenge
causality.
However this is not the case because of the following reason:
The singularity only appears in an infinitely sharp eigenstate of
$p^\mu_1+p^\mu_2$. 
The set up of such a state in principle require so extended apparatus
that one can no longer test causality.
In fact, suppose that we set $p^\mu_1+p^\mu_2$ with a finite spread
$\Delta(\stackrel{\rightharpoonup}{p_1}+\stackrel{\rightharpoonup}{p_2})$
then apparatus has to be no more extended than 
$\frac{\hbar}{\Delta(\stackrel{\rightharpoonup}{p_1}+
		\stackrel{\rightharpoonup}{p_2})}$ 
in $\stackrel{\rightharpoonup}{x}$-space.
But in this case the singularity also gets washed out so as to go into 
the complex plane.

However, even if this is the case, it is unavoidable that the
distances
$r_{ik}=
	\sqrt{(\stackrel{\rightharpoonup}{x_i}-
		\stackrel{\rightharpoonup}{x_k})^2}$
can go to zero even if 
$\stackrel{\rightharpoonup}{x_i}-\stackrel{\rightharpoonup}{x_k}$
does not go to zero.
Thus there will be branch point singularities.

Next we argue how many sheets there are in the $N$ particle scattering.
The wave functions are prepared and thus whether it is analytic or not
may depend on the preparation.
However even if we prepare an analytic wave function there may be
unavoidable branch-point singularities when the distance $r_{ik}$ goes 
to zero for nonzero complex distance vector.
These unavoidable singularities come in from the interactions because 
the function $\eta_0$ is singular at $r=0$.

If we consider the scattering of two particles, it is described by the 
phase shift.
The wave functions of the scattered particles are described in terms
of 
$\jmath_\ell(kr_{12}){\cal Y}^\ell_m
(\stackrel{\rightharpoonup}{x_1}-\stackrel{\rightharpoonup}{x_2})$
and 
$\eta_\ell(kr_{12}){\cal Y}^\ell_m
(\stackrel{\rightharpoonup}{x_1}-\stackrel{\rightharpoonup}{x_2})$,
where $r_{12}$ is the distance of the two particles 1 and 2 and 
$\stackrel{\rightharpoonup}{x_1}$ , 
$\stackrel{\rightharpoonup}{x_2}$ their positions.
The spherical harmonics ${\cal Y}^\ell_m$ are defined as

\begin{equation}
{\cal Y}^\ell_m(\stackrel{\rightharpoonup}{x_1}-x_2)=r^\ell y^\ell_m
\end{equation}
where $y^\ell_m$ are the usual $r_{12}$-independent spherical
harmonics. 
When there is no scattering, it is expressed by 
$\jmath_\ell{\cal Y}^\ell_m$
and there will be no branch point singularities. 
However if there is scattering, the functions

\begin{equation}
\eta_\ell(k\cdot r_{12}){\cal Y}^\ell_m
\end{equation}
come into the wave function which are odd in $r_{12}$ so that the
branch point singularities appear.
Here $r_{12}$ is a simple square root and thus it has two sheets on 
which the wave functions differ by the sign of 
$\eta_\ell(k\cdot r_{12}){\cal Y}^\ell_m$ .
These parts of the wave functions are given by the interaction as was
discussed in the preceding section.
In particular the switch of the sign can be achieved by keeping the
$\jmath_\ell{\cal Y}^\ell_m$ terms fixed but changing the sign  of the 
interaction, $\lambda\rightarrow -\lambda$.

If we consider $N$-particle system, there are $\frac{1}{2}N(N+1)$ pairs 
of particles of which distance is given by 
$r_{ik}=
	\sqrt{(\stackrel{\rightharpoonup}{x_i}-
		\stackrel{\rightharpoonup}{x_k})^2}$.
Since we consider just two sheets with respect to each of these
$\frac{1}{2}N(N+1)$ pairs, there appear $2^{\frac{1}{2}N(N+1)}$ sheets
altogether.

We consider the case that the incoming waves scatter in all possible
combinations with each other.
Thus a scattered wave  $\eta_{ik}$ of particles $i$ and $k$
changes sign such that

\begin{equation}
\sum_{(i,k)}\eta_{ik}\cdot S_{ik} \ \ \ (i\neq k)
\end{equation}
when going from the original sheet to the one characterized by 
$\{S_{ik}\}$.
Here $S_{ik}$ is understood to be $S_{ik}=0$ for the first sheet with
respect to $r_{ik}$ and $S_{ik}=1$ for the second sheet.
So the piece of the wave function changes as 

\begin{equation}
\psi_{piece} \rightarrow
(-1)^{\sum_{(i,k)}\eta_{ik}S_{ik}}\psi_{piece} \ \ \ .
\end{equation}
Obviously this sign changes separate the $2^{\frac{1}{2}N(N+1)}$
sheets and this is the minimal number of sheets which gives the wave
function with minimal singularities of the branch point.

Typically even if the incoming waves have other singularities, we can
continue to other sheet with respect to the $r_{ik}$ distance as
will be discussed in next section.

\section{The CPT-like theorem}

In this section we first consider how the usual CPT symmetry operation[4]
acts on the naive vacuum world.
As will be explained below the naive vacuum is brought into the
totally different state if the usual CPT transformation is applied and
thus it is spontaneouly broken.
We shall then propose a new CPT-like symmetry which leaves the naive
vacuum invariant.

In order to apply the usual CPT operation, we describe the naive
vacuum from the true vacuum viewpoint.
In the naive vacuum viewpoint there are no particles and/or no holes
in all the single particle states with positive energies as well as
negative ones.
Even if we consider them from the true vacuum point of view there are
no particles in the positive energy states.
However the negative energy states are described as the ones with
minus one hole for bosons and one hole for fermions.
If we use the states with only the positive energies there are minus one
particles for bosons and fermions respectively and therefore one anti-particles. 

The usual CPT operation acts on the naive vacuum described from the true 
vacuum viewpoint in the following way : It brings into the positive
energy states with minus one  boson and one fermion.
These states should be translated back to the ones in terms of the
naive vacuum viewpoint.
All the states with positive energies as well as negative ones are
filled by minus one  boson and one fermions.
These are completely new states relative to the naive vacuum and we
conclude that the usual CPT symmetry is spontaneously broken.

We now ask the question if we could invent a CPT-like
operation for which the naive vacuum does not represent such a
spontaneous breakdown.

We shall indeed answer this question affirmatively :
We propose a new strong reflection in which the inversion of the
operator order contained in the usual strong reflection is excluded.
We then compose it with an operation of analytic continuation which
will be presented below.

At first one might think that a primitive strong reflection on a state
in the naive vacuum world be a good symmetry.
This reflection is defined so as to shift the sign of any Lorentz
tensor by $(-1)^{\rm rank}$ .
Thus the Hamiltonian gets a sign shift $H\rightarrow -H$ under the
primitive strong reflection.
Indeed this expectation works for the free theory.
However, looking at the interaction described by the boundary
condition in eq.(\ref{7.12}), we see that under the time reversal
$t\rightarrow -t$ which is contained in the strong reflection the left 
hand side changes the sign while the right hand side does not.
This means that action of the naive strong reflection on a state in the 
Fock space in the naive vacuum world which obeys the condition 
eq.(\ref{7.12})brings into a state which does not obey this equation.
If we should make the symmetry a good one we might do it so by
changing $\tan\delta\rightarrow -\tan\delta$.
It would be achieved by changing sign on the coupling constant
$\lambda$ according to the formula eq.(\ref{7.30}).
However the coupling constant $\lambda$ is one of the fundamental
quantities and it cannot change sign.
Thus the naive strong reflecting without operator order inversion is not 
a good symmetry for the interaction $H_{int}$.

We may look for a modification of the Fock space state of the model
that could provide a sign shift of the interaction.
By combining the modification with the simple strong reflection we
could get a symmetry of the laws of nature in the naive vacuum picture.

For finding this modification it would provide an important hint that
in the relation eq.(\ref{7.12}) there is a factor $\frac{1}{r_{ik}}$ and that
the distance $r_{ik}$ is a square root
$r_{ik}=
\sqrt{(\stackrel{\rightharpoonup}{x_i}-
	\stackrel{\rightharpoonup}{x_k})^2}$.
This distance has a branch point at $r_{ik}=0$ and $\infty$ as an
analytic complex function.
In the analytic continuation it has two sheets on which $r_{ik}$ takes 
the opposite values

\begin{equation}
r_{ik}|_{\rm one \ sheet} = -r_{ik}|_{\rm the \ other \ sheet} \ \ \ .
\end{equation}

This sign change is just what we need, provided we postulate that 
our new CPT-like operation should contain the analytic continuation
onto the opposite sheet in addition to the simple strong reflection.
By the former change we get $r_{ik}\rightarrow -r_{ik}$ and it makes
the boundary condition eq.(\ref{7.12}) turn out to be invariant under the full
operation.
Thus the new CPT-like operation consists of the following two operations;
A) simple strong reflection and B) $r_{ik}\rightarrow
-r_{ik}$ provided by analytic continuation.

In order to get the sign shift $r_{ik}\rightarrow-r_{ik}$ for the
interactions between all possible pairs of particles in the naive
vacuum world it is needed to make the analytic continuation so that it
goes to the opposite sheet with respect to every pair of particles
$i,k$.
In fact we restrict our attention to such a wave function that it has
only the branch points due to the square roots in $r_{ik}$.
With such an assumption for the wave functions we may define a
procedure B) by which a wave function $\psi$ is replaced by a new one
for the same values of 
$(\stackrel{\rightharpoonup}{x_1},
\stackrel{\rightharpoonup}{x_2},\ldots,
\stackrel{\rightharpoonup}{x_N}))$
but on the different sheet.

All this means that we propose a replacement for the usual CPT-theorem 
in order to avoid spontaneous breakdown on the naive vacuum.

We consider the CPT-like theorem
by taking the simple case of $(\varphi^+\varphi)^2$ theory.
It should be valid under very broad conditions.

In the $(\varphi^+\varphi)^2$ theory in the naive vacuum world there
is a symmetry under the following combined operation of A) and B) :

A) Strong reflection without operator inversion.\\
We make a replacement in the wave function of all position variables 
$\stackrel{\rightharpoonup}{x_1}\stackrel{\rightharpoonup}{x_2},\ldots,\stackrel{\rightharpoonup}{x_N}$ by their opposite
ones.

\begin{equation}
\psi_{\pi\varphi\ldots}
	(\stackrel{\rightharpoonup}{x_1},\ldots,
	 \stackrel{\rightharpoonup}{x_N})\rightarrow
	(-1)^{\#\pi-{\rm indices}}
\psi_{\pi\varphi\ldots}
	(-\stackrel{\rightharpoonup}{x_1},\ldots,
	 -\stackrel{\rightharpoonup}{x_N})
\end{equation}
whereby replacements $\stackrel{\rightharpoonup}{p_i}\rightarrow
-\stackrel{\rightharpoonup}{p_i}$ are performed.
In this operation energy and thereby time shift sign.
This means that $\psi_{\varphi\ldots\pi\ldots\varphi}$ components get 
a minus sign for each $\pi$

\begin{equation}
\psi_{\varphi\varphi\pi\varphi\ldots\pi_N}
	(\stackrel{\rightharpoonup}{x_1},\ldots,
	 \stackrel{\rightharpoonup}{x_N})\rightarrow 
	(-1)^{\#(\pi-{\rm indices})}
\psi_{\varphi\varphi\pi\varphi\ldots\pi_N}
	(-\stackrel{\rightharpoonup}{x_1},\ldots,
	  \stackrel{\rightharpoonup}{x_N}) \ \ \ . 
\end{equation}
This transformation A) may be described as strong reflection
though without including the operator inversion rule.
It should be noticed that this operation A) does not keep the inner
product invariant.

B) Analytic continuation of the wave function onto the other sheet
on which all the distances $r_{ik}$ shift sign.

The wave function 
$\psi_{\varphi\pi\ldots}
(\stackrel{\rightharpoonup}{x_1},\ldots,\stackrel{\rightharpoonup}{x_N})$ 
is analytically continued onto another sheet over the complex 3N
dimensional configuration space but with the same values of the
arguments
$\stackrel{\rightharpoonup}{x_1},\ldots,\stackrel{\rightharpoonup}{x_N}$ 
.
The continued sheet is the one on which the relative distances 
$r_{ik}=\sqrt{(\stackrel{\rightharpoonup}{x}_i-
	\stackrel{\rightharpoonup}{x}_k)^2}$ have shifted their sign
relative to the first sheet. 

This CPT-like theorem presupposes to choose boundary conditions for long
distances in the following way :

\begin{enumerate}
\renewcommand{\labelenumi}{\arabic{enumi})}
\item We suppose that in the scattering at $t\rightarrow -\infty$ we have 
ingoing waves while at $t\rightarrow\infty$ outgoing waves. 
Here ingoing and outgoing waves are interpreted according to the
directions of velocities but not momenta.
\item For the negative energy pairs we choose, instead of convergent
boundary conditions anticonvergent boundary conditions, which means
that total energy is negative.
Some explanation on this point is in order.
\end{enumerate}

First of all we call attention to the relations between the velocity
$\stackrel{\rightharpoonup}{\upsilon}$ and momentum
$\stackrel{\rightharpoonup}{p}$.
For positive energy particles $\stackrel{\rightharpoonup}{\upsilon}$
and $\stackrel{\rightharpoonup}{p}$ point in the same direction while
for the negative energy particles they are opposite.
Whether the wave is incoming or outgoing is defined by the velocity.

The operation A) brings an extended bound state over a finite region
into a configuration with a finite extension.
If a bound state wave function behaves as $e^{-kr}$, it may blows up
as $e^{kr}$ after the analytic continuation B).
This is an obvious consequence of changing the sign of all the formal
expression 
$r_{ik}=\sqrt{(\stackrel{\rightharpoonup}{x}_i-
		\stackrel{\rightharpoonup}{x}_k)^2}$
for the relative distances.
Then in order to be able to define the condition on the blow up state, 
we have to analytically continue back these distances to the sheet
with no sign shift relative to the original sign.
Thus we can summarize in the following way :
For positive energy we use ordinary bound state condition while for
negative energy the anti-bound condition is required.

It should be remarked that both energy and momentum change the sign,
but not the velocity.

We may list the properties of operations A) and B) : 

Properties of A)

1) transformations of variables

\begin{eqnarray}
&&\stackrel{\rightharpoonup}{x}
\rightarrow -\stackrel{\rightharpoonup}{x} \ \ \ , \ \ \ 
t\rightarrow -t\nonumber\\
&&\stackrel{\rightharpoonup}{p}
\rightarrow-\stackrel{\rightharpoonup}{p} \ \ \ , \ \ \ 
E\rightarrow -E \nonumber\\
&&\stackrel{\rightharpoonup}{\upsilon}\rightarrow 
\stackrel{\rightharpoonup}{\upsilon}\nonumber\\
&&i\rightarrow  i\ ({\rm  no \  complex \ conjugation})\nonumber
\end{eqnarray}

2) boundary conditions\\
$S$-matrix type boundary condition, that is, ingoing at $t=-\infty$
and outgoing at $t=\infty$ are kept fixed.
In terms of velocity,

\begin{eqnarray}
{\rm ingoing \ wave} &\rightarrow& {\rm outgoing \ wave}\nonumber\\
(\stackrel{\rightharpoonup}{x}\cdot
	\stackrel{\rightharpoonup}{\upsilon}<0)&&
(\stackrel{\rightharpoonup}{x}\cdot
	\stackrel{\rightharpoonup}{\upsilon}>0)\nonumber
\end{eqnarray}

3) particle and hole are not exchanged

\begin{equation}
{\rm particle}\rightarrow{\rm particle} \ , \ 
{\rm hole}\rightarrow{\rm hole} \ \ \ .
\nonumber
\end{equation}

Properties of B)

The operation of the analytic continuation in B) is performed on the
maximally analytic wave function
$\psi(\stackrel{\rightharpoonup}{x_1},
\stackrel{\rightharpoonup}{x_2}\ldots,\stackrel{\rightharpoonup}{x_N})$ 
for any fixed number $N$ of particles.
In this way the number of particles $N$ is conserved.

A continuous function 
$\psi(\stackrel{\rightharpoonup}{x_1},
\stackrel{\rightharpoonup}{x_2}\ldots,\stackrel{\rightharpoonup}{x_N})$ 
of variables $(\stackrel{\rightharpoonup}{x_1},
\stackrel{\rightharpoonup}{x_2}\ldots,\stackrel{\rightharpoonup}{x_N})$
conceived of as complex variables is analytically continued along a
path which avoids the branch points.
It leads onto a sheet characterized by the change of the sign for
all the $r_{ik}$ relative to the starting one.

In a typical scattering the scattered wave function will
change sign under the operation.
In the expression in terms of the partial wave of the wave function 

\begin{eqnarray}
&&\varphi_{\varphi\varphi}\propto
	\cos\delta\jmath_0(k\cdot r)-\sin\delta n_0(k\cdot r)  \ \ \ , 
\nonumber\\
&&\varphi_{\varphi\pi}\propto
	\omega\cos\delta\jmath_0(k\cdot r)-
	\omega\sin\delta n_0(k\cdot r)
\end{eqnarray}
we make analytic continuation $r\rightarrow -r$ and 
$n_0 \rightarrow -n_0$ while $\jmath_0\rightarrow \jmath_0$ so that
the scattered wave given by $n_0$ changes sign.

\section{Proof of the CPT-like theorem}

To prove the CPT-like theorem discussed in last section we first
remark that it is true for free theory : 
In this case the operation A) may be brought upon in momentum space and 
both energy and momentum are inverted for all particles.
Since free wave function is analytic and it is the same on all the
sheets.
Thus the continuation operation B) has no effect in the free case.

If there are interactions the strong reflection A) does not keep
invariant the boundary conditions at the meeting point

\begin{equation}
i({\rm coeff. \ of \ }
\varphi_{\varphi\varphi\pi\ldots\varphi}|_{\frac{1}{r}-{\rm part}})=
	2\tan\delta\cdot({\rm coeff. \ of \ }
	\varphi_{\varphi\varphi\pi\ldots\varphi}|_{\rm const.part}) 
	\ \ \ .
\label{10.1}
\end{equation}
with $\tan\delta=-\frac{\lambda}{64\pi}$
because the left hand side changes the sign under $t\rightarrow -t$.
Thus after the operation A) the condition is satisfied with opposite
sign $\tan\delta\rightarrow-\tan\delta$.
On the other hand with analytic continuation under B) the equation of
motion in the $r_{ik}\neq 0$ regions remains to be satisfied with the
eigenvalue $-E$.
However under the operation A) the solution with $E$ becomes the one with 
$-E$.
But on return the $r_{ik}$'s have changed sign.
Thus after both A) and B) operations not only the equation of motion is 
fulfilled but also the boundary condition eq.(\ref{10.1}) is satisfied with
the original correct sign.

Let us remark on the boundary condition at large
$\stackrel{\rightharpoonup}{x_i}$'s.
In this case as long as the scattering is negligible the $\jmath_\ell$ 
expansions work which is even at $r=r_{ik}$
in the sense that if one wants to use spherical harmonics one needs

\begin{equation}
{\cal Y}^m_\ell
(\stackrel{\rightharpoonup}{x_i}-\stackrel{\rightharpoonup}{x_\ell})=r^\ell 
y^m_\ell\left(\frac{\stackrel{\rightharpoonup}{x_i}-
	\stackrel{\rightharpoonup}{x_\ell}}{r}\right) \ \ \ .
\end{equation}
The final expansion turns out even function in $r$ when no $n'_\ell$s are
present.
Thus the wave function is the same one on all sheets and the
operation B) is unaffected.
When interaction is switched on in an incoming wave, the
sign is changed for a piece of scattered wave when scattered an odd 
number of times. 
However, the form of the waves, either 
$e^{ik|\stackrel{\rightharpoonup}{x}|}$ or 
$e^{-ik|\stackrel{\rightharpoonup}{x}|}$, is not changed by the analytic
continuation  B).
Thus the velocity-wise boundary condition can be settled as if there
were only the operation A) in the CPT-like theorem.
Under A) we have the shifts 
$\stackrel{\rightharpoonup}{x}\rightarrow
-\stackrel{\rightharpoonup}{x} \ , \ 
\stackrel{\rightharpoonup}{p}\rightarrow
-\stackrel{\rightharpoonup}{p} \ , \ 
E\rightarrow -E$ and thus 
$\stackrel{\rightharpoonup}{\upsilon}=
	\frac{\stackrel{\rightharpoonup}{p}}{E}$
is unchanged.
Therefore 
$\stackrel{\rightharpoonup}{x}\cdot
\stackrel{\rightharpoonup}{\upsilon}\rightarrow-
\stackrel{\rightharpoonup}{x}\cdot
\stackrel{\rightharpoonup}{\upsilon}$.

In the $S$-matrix conditions the ingoing wave packets at $t=-\infty$
have
$\stackrel{\rightharpoonup}{x}\cdot\stackrel{\rightharpoonup}{\upsilon}<0$ 
while the outgoing ones have 
$\stackrel{\rightharpoonup}{x}\cdot\stackrel{\rightharpoonup}{\upsilon}>0$.
Combining with $t\rightarrow -t$, this condition remains the same
under A) and thus under the whole CPT-like operation is unchanged in the naive
vacuum world.

To suggest that this CPT-like theorem is very general we sketch the
method how to construct a general proof.
First operation A) is in fact strong reflection but without the
operation of order change of the creation and annihilation operators.
It is strongly suggested to perform an analytic continuation of a
Lorentz-boost to an imaginary rapidity

\begin{equation}
\eta=i\pi
\end{equation}
where the boost velocity is given by 
$\upsilon_{boost}=\tanh \eta$.
It is followed by a $\pi$ rotation around the boost axis.

It may be written, for simplicity, in terms of the matrix elements of
the Lorentz transform as 

\begin{equation}
\Lambda^\mu\nu=\left(
\begin{array}{cccc}
\cosh\eta&\sinh\eta&0&0\\
-\sinh\eta&\cosh\eta&0&0\\
0&0&1&0\\
0&0&0&1\\
\end{array}\right) \ \ \ .
\end{equation}
However the situation is not as simple as this.
In order to perform a Lorentz-boost of a wave function we have to
live with the problem that simultaneity is not an absolute concept.
To obtain the wave function in another reference frame, it is required 
that one time develops the state of different positions in a different 
way.
The best way may be to use a formalism in which different times are
allowed for each particle.
Such a formalism is possible locally since the interactions are so
local that the theory is free and we can time develop one particle but 
not others.

In the neighborhood of
$(\stackrel{\rightharpoonup}{x_1},\ldots,\stackrel{\rightharpoonup}{x_N})$ 
for $N$ particles with
$\stackrel{\rightharpoonup}{x_i}\neq\stackrel{\rightharpoonup}{x_\jmath}$ 
we may treat them as free particles for a certain finite time.
Then up to the final time we consider that each particles have
individual times $t_1,\ldots,t_N$.
It may be allowed to boost them with small $\eta$.
For boosts with large $\eta$ the Lorentz transformations will involve
coinciding $\stackrel{\rightharpoonup}{x_k}$'s so that the interaction 
caused by singularities will be associated.

We would like to suggest that under a properly performed Lorentz-boost 
with the imaginary rapidity $\eta=i\pi$ we may be able to let all the
distances 
$r_{ik}=
\sqrt{(\stackrel{\rightharpoonup}{x_i}-
\stackrel{\rightharpoonup}{x_k})^2}$
go onto the second sheet.
This means that the analytic continuation of Lorentz boosts and $\pi$
rotation leads not only to the A) operation but also to B).
Thus our CPT-like theorem turns out to be the analytic continuation of 
the Lorentz transformations.

To see typical sheet structure we consider the two particle case in which
they move freely in the classical approximation and pass each other at 
some finite distance or meet at a point.
For simplicity we take a situation in which one of the particles
passes through the origo of the space time coordinate system.
When making a pure Lorentz transformation this particle will continue
to cut the $t'=0$ simultaneity hyperplane in the same event.
This means that the distance between the passage events on the
simultaneity hyperplane of the two particles is given for all
reference frames by the one from the origo to the passage event of
number two particle.

For simplicity we shall also choose the initial coordinate system in
such a way that the particle number two is at rest.
Then the distance between the two particles at zero time $t'=0$ in the 
boosted coordinate system is equal to the Lorentz contracted length
of the stick of which one end is at the origo in the original frame.
Let us suppose that the stick relative to the $x$-axis in the original 
frame have the longitudinal length component $\ell_{''}$ and the
transverse one $\ell_\bot$.
The coordinates of the three space in the original coordinate system
of the particle 2 passage read

\begin{equation}
\stackrel{\rightharpoonup}{x}=
	(\ell_{''} \ , \ \ell_\bot \ , \ 0)
\end{equation}
by choosing that the $x-y$ plane contains the stick.
The original length from the point $\stackrel{\rightharpoonup}{x}=0$
to $\stackrel{\rightharpoonup}{x}=(\ell_{''} \ , \ \ell_\bot \ , \ 0)$ 
is given by $\ell=\sqrt{\ell^2_{''}+\ell^2_\bot}$ and gets Lorentz
contracted to be 

\begin{equation}
\ell_{(\rm new \ frame)}=\sqrt{\ell^2_\bot+\ell^2_{''}(1-\upsilon^2)}
\end{equation}
where $\upsilon$ is the boost velocity along the $x$-axis in the new
frame. 
In terms of rapidity it reads

\begin{equation}
\ell_{(\rm new \ frame)}=\sqrt{\ell^2_\bot+\ell^2_{''}(1-\tanh^2\eta) }
\end{equation}
This can be written as the factorized radicant form

\begin{equation}
\ell_{(\rm new \ frame)}=\sqrt{(\ell_\bot -i\frac{\ell_{''}}{\cosh\eta})
	(\ell_\bot +i\frac{\ell_{''}}{\cosh\eta})} \ \ \ .
\end{equation}
This reveals the branch points for $\ell_{(\rm new \ frame)}$as a
function of the boosting rapidity $\eta$

\begin{eqnarray}
\eta&=&\makebox{arcosh}\left(\pm\frac{i\ell_{''}}{\ell_\bot}\right) =
i\frac{\pi}{2}\pm\makebox{arcsinh}\left(\ell_{''}/{\ell_\bot}\right) \ \ \ ,\\
\makebox{and }&&\nonumber\\
\eta&=&\makebox{arcosh}(0)=i\frac{\pi}{2}(\makebox{modi}\pi) \ \ \ .
\end{eqnarray}
The result of a boost by $\eta=i\pi$ may depend on which path of
$\eta$ from 0 to $i\pi$ one chooses.
Notice that we cannot let $\eta$ go along the pure imaginary axis
because it would hit the two branch points at $\eta=i\frac{\pi}{2}$ .
In fact each of the two factors 
$\sqrt{\ell_\bot\pm i\frac{\ell_{''}}{\cosh\eta}}$ 
has a branch point at $\eta=i\frac{\pi}{2}$.
Thus there is only a pole and no branch point for the $\ell_{(\rm new
\ frame)}$ expression.
Therefore it does not matter to the endresult of $\ell_{(\rm new \
frame)}$ on which side of the  singularity at $\eta=i\frac{\pi}{2}$
the passage has been performed to reach the endpoint $\eta=i\pi$.
So we can decide to make the path go by deforming to either side by
infinitesimal amounts $\pm\epsilon$ around this pole and the end
result will be the same.
We take the path outside the branch point singularities at 
$\eta=i\frac{\pi}{2}\pm \sinh^{-1}\frac{\ell_{''}}{\ell_\bot}$.
But these are at finite distance from the imaginary axis as long as
$\ell_{''}\neq 0$.
In this way we have seen that for $\ell_{''}\neq 0$ the most direct
path is the one connecting those two true branch points which is
almost the imaginary axis except for $\pm\epsilon$ deformation.
If we choose this path it is easy to show that a sign of
$\ell_{(\rm new \ frame)}$ gets changed in the sense that 

\begin{eqnarray}
\ell_{(\rm new \ frame)}(\eta=i\pi)&=&
	-\ell_{(\rm new \ frame)} (\eta=0)\nonumber\\
	&=&-\ell
\end{eqnarray}
Since all 4 vectors change sign by going to $\eta=i\pi$ the stick
should lie between $\stackrel{\rightharpoonup}{x}=0$ and
$\stackrel{\rightharpoonup}{x}=(-\ell_{''} \ , \ -\ell_\bot \ , 0)$ .
If we write pure imaginary $\eta$ as $\eta=i\hat\eta$ where $\hat\eta$ 
is real, we have $\cosh\eta=\cos\hat\eta$.
Thus the full radicant

\begin{equation}
\ell^2_\bot+\frac{\ell^2_{''}}{\cosh^2\eta}=
	\ell^2_\bot+\frac{\ell^2_{''}}{\cos^2\hat\eta}
\end{equation}
is positive and lies on the purely imaginary axis with
$\eta=i\hat\eta$.
If the radicant remains real and nonzero 
the square root $\eta_{(\rm new \ frame)}$ would stay positive say, by 
continuity.
Only when we make the detour with $\pm\epsilon$ is there a possibility 
to let $\eta_{(\rm new \ frame)}$ go into the complex quantity and
thus it may possibly change sign
when returning to the imaginary axis.

In fact since there is the pole in 
$\ell_{(\rm new frame)}$ at $\eta=i\frac{\pi}{2}$ the length
$\ell_{(\rm new \ frame)}$ changes sign when $\eta$ passes
from 
$\eta=i\left(\frac{\pi}{2}-\hat\epsilon\right)$ to
$\eta=i\left(\frac{\pi}{2}+\hat\epsilon\right)$ along a little detour
away from the imaginary axis.
Thus the result is that along the path described above there is indeed 
a sign change on the $\ell_{(\rm new \ frame)}$.

This means that for other pairs of particles we could also get such a
sign shift for the distance between the particles $r_{ik}=\ell_{(\rm new \ 
frame)}$.
We then could achieve both the operations A) and B) as an analytic
continuation of a Lorentz transformation.

We will not prove here the fact that one can always find a path along
which the sign shifts on such distance.
However, we showed that by choosing as the suggestive shortest path of 
analytic continuation the one along the imaginary $\eta$-axis we get
the sign shift. 

So if it should at all be possible not to switch the signs of
distances by making an analytic continuation to $\eta=i\pi$, it would
at least be a more complicated and/or far away path.


\section{Relation to the usual CPT-theorem}

As we have described in the previous section we think that the new
CPT-like theorem holds in the naive vacuum world where there are a
finite number of particles.
However, if we are not afraid of infinities we may apply it on the
correct vacuum where all the fermion negative energy states are filled 
and for the bosons one particle is removed from each negative energy
single particle states.
In this case, however, the state is brought into another world in
which the positive energy states are all filled for fermions and for
bosons one particle is removed from each positive energy single
particle states.
Thus it is not a practical symmetry for the real world described by
the correct vacuum.
The CPT-like symmetry brings the real world across the figure with the 
four vacua as is illustrated in Fig.1, in much the same way as the
usual CPT-theorem brings the naive vacuum across the figure.

If we take a point of view that we apply both new and usual CPT-theorems 
on all the four vacua and associated worlds, we may consider the
composite operation ;
\begin{eqnarray}
&&\makebox{(CPT-like \ symmetry)}\cdot\makebox{(usual \ CPT symmetry)}\nonumber\\
&=& 
\makebox{particle-hole \ exchange \ with \ analytic \ continuation  }
\nonumber
\end{eqnarray}
In fact the composite operation is the symmetry under the following
combined set of three operations :

\begin{enumerate}
\renewcommand{\labelenumi}{\arabic{enumi})}
\item  Particle hole exchange ; it replaces every particles by
corresponding holes which is equivalent to the $-1$ particle in the
same single particle state.
\item  Perform the same analytic continuation to all the particles so as
to bring all the relative distances $r_{ik}$ of them to change sign
\item  Complex conjugation of the wave function.
This operation is a symmetry in the sense that it brings the energy and 
momentum to the ones with opposite sign.
\end{enumerate}

It should be noted that this composite operation is antiunitary.
This symmetry can be derived under very general conditions that
consist mainly of the locality of interactions and some remnant of
Lorentz symmetry.
The point is that in order to prove the sign change of the
interactions by going to the other sheet in the distances
$r_{ik}\dot\psi$ in the boundary conditions eq.(\ref{10.1}) should be shown to
behave as an odd power of ${r_{ik}}^{-1}$.

Certainly one can claim that the particle hole exchange theorem may be 
proven by combining our arguments in section 5 for the proof of the
CPT-like theorem in the naive vacuum with the usual CPT theorem.
However we believe that the proof could be made under much weaker
assumptions.

\section{Conclusion}
We have investigated some properties of a spinless boson theory with a
$|\mathbf{\Psi}|^{4}$ term conserving the particle number and found that the
few particle state with incoming momentum eigen-state particles
become in reality described by in principle explicitly calculable wave
functions for these few particles.
They even have nice analytical properties then with a controbable
number of sheets $2^{\frac{1}{2}N(N+1)}$.
The main result is then to formulate and prove a theorem reminiscent
of the CPT-theorem but now for a world 
in the playground state of having just a finite number of
particles inserted extra relative to the naive vacuum (i.e. naively
second quantized theory with just ``a few'' particles present).
This CPT-like theorem means that there is then a symmetry under an
operation composed of two operation A) and B).
Here A) is the usual strong reflection without though having operator
inversion.
I.e. the order of operators are kept in A).
The operation B) is an analytic continuation to a certain other sheet
of the wave function.
Really it is to the sheet on which all the inter-particle distances
$r_{ij}$ have changed sign. 

This kind of rules for the pre-filled world picture is from which one
might put in perspective the usual and correct theory.

Our study of the situation with Dirac seas that are not yet filled
is first of all a mathematical game meant for seeking a perspective as
dramatic as the Dirac sea for fermions.  It is therefore to study the
mathematical simplifications or modified rules -- such as our above
attempt on a modified CPT - like theorem, which are the sort of things
one could hope to get out of the Dirac sea method of looking at
the boson quantization.

A very promising direction for use of the Dirac sea for bosons method
is to use it for putting supersymmetry in even at the single particle
level, which is without such a procedure as ours impossible.  We have
already ref[5] on this subject, but we propose to more direct
formulation of SUSY at the single particle level in the forthcoming paper[6].

\section*{Acknowledgement}
We would like to acknowledge Y. Habara for useful discussion on
supersymmetry. Main part of this research was carried out at YITP
while one of the authors(H.B.N.) stayed there as a visiting
professor. H.B.N. is grateful to YITP for hospitality during his
stay. M.N. acknowledges N.B.I. for hospitality extended to him during
his visit.  The work is partially supported by Grants-in-Aid for
Scientific Research on Priority Areas, Number of Areas 763, ``Dynamics
of Strings and Fields'', from the Ministry of Education of Culture,
Sports, Science and Technology, Japan.


\end{document}